\documentclass[preprintnumbers, twocolumn, floatfix,preprintnumbers,showpacs,
letterpaper,
superscriptaddress,nofootinbib]{revtex4}
\usepackage{graphicx,epsfig}
\usepackage{amsmath}
\usepackage {amssymb}
\usepackage{subfigure}
\usepackage{color}

\begin{document}

\title{Emergent universe in theories with natural UV cutoffs}

\author {Mohsen Khodadi}
\email{m.khodadi@stu.umz.ac.ir}
\affiliation{Department of Physics, Faculty of Basic Sciences,\\
University of Mazandaran, P. O. Box 47416-95447, Babolsar, Iran}

\author {Kourosh Nozari}
\email{knozari@umz.ac.ir}
\affiliation{Department of Physics, Faculty of Basic Sciences,\\
University of Mazandaran, P. O. Box 47416-95447, Babolsar, Iran}
\affiliation{Center for Excellence in Astronomy and Astrophysics (CEAAI-RIAAM),\\
P. O. Box 55134-441, Maragha, Iran}

\author{Emmanuel N. Saridakis}
\email{Emmanuel_Saridakis@baylor.edu}

\affiliation{Chongqing University of Posts \& Telecommunications, Chongqing, 
400065,China}

\affiliation{Department of
Physics, National Technical University of Athens, Zografou Campus
GR 157 73, Athens, Greece}
\affiliation{CASPER, Physics Department, Baylor University, Waco, TX 76798-7310, USA}

\begin{abstract}

We investigate the realization of the emergent universe scenario in theories with natural
UV cutoffs, namely a minimum length and a maximum momentum, quantified by a new
deformation parameter in the generalized uncertainty principle. We extract the
Einstein static universe solutions and we examine their stability through a phase-space
analysis. As we show, the role of the new deformation parameter is crucial in a
twofold way: Firstly, it leads to the appearance of new Einstein static universe critical
points, that are absent in standard cosmology. Secondly, it provides a way for a 
graceful exit from the Einstein static universe into
the expanding thermal history, that is needed for a complete and successful realization of
the
emergent universe scenario.  
\end{abstract}


\maketitle

\section{Introduction}

According to the concordance model of cosmology our universe has most probably begun from
an initial singularity at a finite past. The introduction of the inflation paradigm as a
successful way to solve the  horizon, flatness and magnetic monopole problems
\cite{Guth:1980zm}, did not affect the initial singularity issue, which is still
considered as a potential, conceptual disadvantage \cite{Borde:2001nh}.
Finally, in order to describe the observed
late-time universe acceleration a cosmological constant was added, leading eventually to
the $\Lambda$CDM cosmology, namely   the Standard Model of the universe. Nevertheless,
in spite of the remarkable successes of this paradigm, its physical content relating to
the two accelerating phases at early and late times is still not satisfactory, and
furthermore, the initial singularity problem remains open.

There are two ways one could follow in order to bypass the initial singularity problem.
The first is to consider the scenario of bouncing cosmology, in which the current
universe expansion followed a previous contracting phase, with the scale factor being
always non-zero \cite{Mukhanov:1991zn,Novello:2008ra}. The second is to consider  the
scenario of ``emergent universe''  \cite{Ellis:2002we}, in which the universe originates
from
a static state, namely from the ``Einstein static universe'', and then it enters
the inflationary phase, without passing from any singularity. However, both these
alternative cosmological scenarios cannot be obtained in the framework of general
relativity. Concerning the Einstein static universe, which is a necessary ingredient of
the emergent universe
scenario, it can be shown that it is significantly affected by the initial conditions
such as perturbations, which dominate at the Ultra-Violet (UV) limit, and hence it is
indeed unstable against classical perturbations which eventually lead it to collapse
to a singularity \cite{Eddington}.

In order to alleviate the above problems one may follow the way to introduce new
degree(s) of freedom, beyond the standard model of particle physics or/and general
relativity. A first direction is to consider exotic forms of matter that could provide a
successful description of the universe behavior in the framework of general relativity
(see \cite{Copeland:2006wr,Cai:2009zp}  and references therein). The second direction is
to construct a gravitational modification whose extra degrees of freedom could describe
the universe at large scales, while still possessing general relativity as a particular
limit \cite{Nojiri:2006ri,Capozziello:2011et}. Concerning the initial singularity issue,
modified gravity, amongst others, can trigger the cosmological bounce
\cite{Brandenberger:1993ef}, or it can  cure the emergent universe scenario by making
Einstein static universe stable. In particular, the Einstein static universe and thus the
emergent universe scenario, can be successfully realized in various gravitational
modifications, such
as in Einstein-Cartan theory \cite{Boehmer:2003iv}, in $f(R)$ gravity
\cite{Barrow:1983rx}, in $f(T)$ gravity
\cite{Wu:2011xa}, in loop quantum cosmology \cite{Mulryne:2005ef}, in massive gravity
\cite{Parisi:2012cg}, in Ho\v{r}ava-Lifshitz gravity \cite{Khodadi:2015fav}, in
braneworld models \cite{Gergely:2001tn,Atazadeh:2014xsa}, in null-energy-condition
violated theories
\cite{Cai:2012yf} etc, although the successful exit from the
Einstein static universe towards the subsequent expanding thermal history is not always
achieved.

One interesting gravitational modification in the UV regime arises through the use of the
``generalized uncertainty principle'' \cite{Hossenfelder:2012jw}, which seen as a quantum
gravity approach might be related with other quantum gravity  models such as Double
Special Relativity \cite{AmelinoCamelia:2000mn} and string theory \cite{Amati:1988tn}.
Although one can have more than one generalizations of the uncertainty principle, the most
interesting one is when one modifies the standard Heisenberg algebra by a linear and a
quadratic term in Planck length and momentum respectively, which leads to the existence of
two natural UV cutoffs, namely minimum length and maximum momentum
\cite{Ali:2009zq}. Hence, when applied to a cosmological framework, these natural cutoffs
give rise to extra terms in the Friedmann equations, which can have interesting
implications.

In the present work we are interested in investigating the Einstein static universe and
the emergent universe scenario in the framework of theories with natural UV cutoffs. In
particular, we show how the induced extra terms in the cosmological equations lead to
the realization and stability of the Einstein static universe, as well as offering the
mechanism to a phase transition to the inflationary era and the subsequent thermal
history of the universe.

The plan of the manuscript is the following: In section \ref{model}, we briefly review
generalized uncertainty principle with two UV cutoffs, and we apply it in a
cosmological framework.  In section
\ref{dynamicalanalysis} we extract the Einstein static universe solutions. In section
\ref{dynamicalanalysis2} we examine the stability of
the Einstein static universe by performing a dynamical system analysis, studying its exit
towards the inflationary era.   Finally, in
section \ref{Conclusions} we provide the conclusions.

\section{Theories with natural UV cutoffs and their cosmology}
\label{model}

In this section we briefly present theories with natural UV cutoffs, and then we apply
them
in a cosmological framework. As we mentioned in the Introduction, in general this kind
of theories arise from the consideration of generalizations of the uncertainty
principle \cite{Hossenfelder:2012jw}. Although one may have more than one such
generalizations, in the present work we focus on the generalization with two natural
UV cutoffs, namely a minimum length and a maximum momentum \cite{Ali:2009zq}. In order to
achieve this, one starts by modifying the standard Heisenberg algebra at high energy
scales, by a linear and a quadratic term in Planck length and momentum respectively, as
\begin{equation}
\label{Int01}
[x_i,p_j] = i \hbar \left[\delta_{i j} - \alpha \left(p
\delta_{i j}\! +\! \frac{p_i p_j}{p}\right) +  \alpha^2 \left(p^2
\delta_{i j}\! + \! 3 p_i p_j\right) \right],
\end{equation}
with $i,j=1,2,3$, where via the Jacobi identity \cite{Ali:2011fa} it is guaranteed that
$[x_i,x_j]=0=[p_i,p_j]$. The parameter $\alpha$ quantifies the quantum gravity
deformation parameter, and can alternatively be written as
$\alpha=\frac{\tilde{\alpha}}{cM_{pl}}=\frac{\ell_{pl}}{\hbar}\tilde{\alpha}$, with
$M_{pl}$ and $\ell_{pl}$ the Planck mass and Planck length respectively, $c$ the speed of
light, and $\hbar$ the induced Planck constant. The
dimensionless parameter $\tilde{\alpha}$ according to experiments is bound to be smaller
than $10^{11}$, however theoretical arguments suggest that its value should be around 1,
in order for minimal length effects to be important only around the Planck length and not
introduce a new physical scale between the Planck and the electroweak scale
\cite{Ali:2011fa,Marin:2013pga,Das:2008kaa}.

Let us now apply the above generalized uncertainty principle in a cosmological framework.
Since the quantum gravity deformation parameter $\alpha$ is expected to have effects only
at high energy scales, we will focus on the early-time phases of the cosmological
evolution, which indeed will correspond to the realization of the emergent universe.
We start by considering the homogeneous and isotropic Friedmann-Robertson-Walker (FRW)
geometry, with metric
\begin{equation}
\label{2-0}
ds^2=- N^2 dt^2+a^2(t) \left(\frac{dr^2}{1-k~r^2}+r^2 d\Omega^2\right),
\end{equation}
where $a(t)$ is the scale factor and $N$ the lapse function, and with
$k=0,+1,-1$ corresponding to flat, close and open spatial geometry respectively.

One can extract the field equations in the above metric, i.e. the Friedmann equations,
via the Hamiltonian constraint $\mathcal{H}_E=0$, namely \cite{Ali:2014hma}:
\begin{equation}
\label{2-01}
\mathcal{H}_E=\frac{\kappa}{4} \frac{N p_a^2}{a} + \frac{N a k}{\kappa} - N a^3
\rho + \lambda \mathcal{P},
\end{equation}
with $\kappa\equiv1/3M_{pl}^2=8\pi G/3$ the gravitational constant, and
where $\lambda$ and $\mathcal{P}$ are the Lagrange multiplier and the momentum conjugate
to the lapse function $N$, respectively. In the above expression
$\rho$ is the energy density of the universe content, corresponding to a perfect fluid
with equation-of-state parameter $w$.

In general, for two typical variables $A$ and $B$, the Poisson brackets are defined as $
\{A,B\}=\left(\frac{\partial A}{\partial
x_i}\frac{\partial B}
{\partial p_j}- \frac{\partial A}{\partial p_i}\frac{\partial B}{\partial x_j}\right)
\{x_i,p_j\}$, where the  canonical variables $x_i$ and $p_j$ in the cosmological context
are replaced by $a$ and $p_a$, respectively. Although using the standard uncertainty
principle they satisfy the usual relation $\{a,p_a\}=1$, considering the deformed Poisson
algebra that arises from the generalized uncertainty principle (\ref{Int01}), up to
first order in $\alpha$, the Poisson bracket between $a$ and $p_a$ becomes
\cite{Ali:2009zq}
\begin{equation}\label{2-02}
\{a,p_a\}= 1-2 \alpha p_a\,.
\end{equation}
Hence, using the Poisson algebra
we obtain the following modified equations of motion
\begin{eqnarray}\label{2-03}
&&
\dot{a}= \{a,\mathcal{H}_E\}= \frac{\partial \mathcal{H}_E}{\partial p_a} (1-2 \alpha
p_a),\\
&&\dot{p}_a=\{p_a,\mathcal{H}_E\}= - \frac{\partial \mathcal{H}_E}{\partial a} (1-2 \alpha
p_a).
\end{eqnarray}
Inserting $\mathcal{H}_E$ from (\ref{2-01})  in the above equations,
using its constraint value $\mathcal{H}_E=0$, and combining them, we finally extract the
first Friedmann equation, namely
\footnote{In principle, the modified Friedmann equations should be derived from a full
quantum gravitational action corresponding to the fundamental theory. However, since
such a quantum-gravity modified action is still unknown, it is common in literature
to consider quantum-gravitational effects phenomenologically, i.e. by deforming the
standard commutation relations as has been mentioned above.
Nevertheless, we mention that there is an inverse method to generate a canonical
Hamiltonian structure,
and subsequently an action, from arbitrary modifications of the dynamical equations,
as it was formulated in detail in  \cite{Singh:2015jus}.}
\begin{equation}\label{2-1}
\bigg(\frac{\dot{a}}{a}\bigg)^{2}=\kappa\rho-\frac{kc^{2}}{a^{2}}-2\sqrt{2\kappa}
\alpha c^{2} a^{2}\rho^{3/2}\bigg(1-\frac{kc^{2}}{\kappa a^{2}\rho}\bigg)^{3/2}~.
\end{equation}
Additionally, taking the time-derivative of this equation, and using also the usual
energy conservation relation
\begin{equation}\label{2-2}
\dot{\rho}+3\frac{\dot{a}}{a}\rho(1+w)=0~,
\end{equation}
we arrive at the second Friedmann equation, namely
\begin{eqnarray}
\label{2-3}
&&\!\!\!\!\!\!\!\!\!\!\!\!\!\!
\frac{\ddot{a}}{a}=-\frac{\kappa}{2}(1+3w)\rho-7\sqrt{\frac{\kappa}{2}}
\alpha c^{2} a^{2}\rho^{3/2}\left(1-\frac{kc^{2}}{\kappa a^{2}\rho}\right)^{3/2}
\nonumber\\
&&\!\!\!\!\!\!\!\!\!\!
+
3\alpha c^{2} a^{2}\rho^{1/2}\!\left[\sqrt{\frac{\kappa}{2}}(1\!+\! 3w)\rho\!-\!\sqrt
{\frac{2}{\kappa}}\frac{kc^{2}}{ a^{2}}\right]\!\!\left(\!1\!-\!\frac{kc^{2}}{\kappa
a^{2}\rho}\right)^{\!1/2}\!\!\!.
\end{eqnarray}
As we observe, the two Friedmann equations (\ref{2-1}) and  (\ref{2-3}) include
terms with the quantum gravity deformation parameter $\alpha$, i.e  they have been
modified by the generalized uncertainty principle. As expected, in the limit
$\alpha\rightarrow0$ they give rise to the standard Friedmann equations.
 It is necessary to note that similar UV modified Friedman
equations
have been discussed earlier in the context
of cosmology induced from other quantum gravity approaches such as loop quantum
gravity (LQG) \footnote{ In Refs. \cite{Ashtekar:2003,
Hossain:2010}
it has been shown that the generalized
uncertainty principle  can be deduced in the context of LQG due to polymer
quantization of
the background spacetime geometry. Hence, one could consider that the
corrections appearing in the
dynamical equations, might arise from LQG with some higher spin representation.}
and Snyder noncommutative geometry, see Refs. \cite{Ashtekar:2006es, Singh:2010qa}
and \cite{Battisti:2009zzb}.

 Finally, as we have mentioned, we note that the
above modified Friedmann equations have been extracted keeping terms up to first order in
$\alpha$. If we additionally keep terms up to second order in $\alpha$
in the deformed Poisson bracket (\ref{2-02}), the corresponding modified Friedmann
equations will include terms such as $\alpha^2 a^4\rho^2\left(1-\frac{kc^{2}}{\kappa
a^{2}\rho}\right)^{2}$. Although mathematically these extra terms will have an effect on
the existence and stability of the critical points that will be analyzed in the
following, in the energy scales that are of interest in the present work, namely those
that correspond to pre-inflation epoch where $\rho\gg\frac{|k|c^{2}}{\kappa}$, the
effects of $\alpha^2$-dependent terms are very tiny and negligible in
comparison with $\alpha$-dependent terms. Moreover, we notice that the
commutator relation (\ref{Int01}), even in the absence of $\alpha^2$ term matches with
other models of generalized
uncertainty principle  and well-known approaches
to quantum gravity, string theory and doubly spacial relativity, see for instance
\cite{Ali:2009zq}. Hence, in summary, the first-order-in-$\alpha$ approximation is very
efficient, and can capture the main effects of natural UV cutoffs.

\section{Einstein static universe}
\label{dynamicalanalysis}

In this section we show that in the cosmological application of the generalized
uncertainty principle the Einstein static universe can be realized.
Let us extract the Einstein static universe solutions.  Inserting the conditions of
the Einstein static universe, i.e.  a
constant scale factor $a=a_{s}$, with $\ddot{a}|_{a=a_{s}}=\dot{a}|_{a=a_{s}}=0$, at an
energy density $\rho=\rho_s$, in the two Friedmann equations (\ref{2-1}) and
(\ref{2-3}), and focusing for mathematical convenience (although this is not necessary)
on the regime $\rho\gg\frac{|k|c^{2}}{\kappa}$ (which is a very robust approximation
since
in SI units it becomes $\rho\gg|k|\times10^{27}$ kg m$^{-3}$, while we know that the
energy
density corresponding to (pre)-inflation scale is $\sim10^{93}$ kg m$^{-3}$)
we find
\begin{equation}
\label{3-3}
\kappa\rho_{s}-\frac{kc^{2}}{a_{s}^{2}}+\frac{6\alpha kc^{2}}{\sqrt{2\kappa}}
\rho_{s}^{1/2}-2\sqrt{2\kappa}\alpha c^{2} a_{s}^{2}\rho_{s}^{3/2}=0~,
\end{equation}
\begin{eqnarray}
&&\label{3-4}
\frac{3k^{2}c^{4}\alpha}{\sqrt{2\kappa^3\rho_{s}}} \frac{1}{a_{s}^{4}} +
\left[\frac{3kc^2}{2\sqrt{2\kappa}}(2-w) \alpha -\frac{\kappa}{2}(1+3w)
\rho_{s}\right]
\frac{1}{a_{s}^{2}}\nonumber\\
&&\ \ \ \ \ \ \ \ \ \ \ \ \ \ \ \
-2\sqrt{2\kappa}\alpha \rho_{s}^{3/2}=0~.
\end{eqnarray}
The solution of this system of algebraic equations will give the critical points of the
cosmological scenario at hand, namely the pair of values for $\{a_s,\rho_s\}$ that
correspond to Einstein static universe solutions. For convenience we study the flat and
non-flat cases separately.

\subsection{Flat universe ($k=0$)}
\label{dynanalflat}

In the case of a flat geometry, and for a general $w\neq-1$, the system
(\ref{3-3}),(\ref{3-4})
accepts only the trivial solution $a_s\rightarrow\infty, \rho_s\rightarrow0$,
independently of the values of $\alpha$. However, in the special case where $w=-1$, i.e
where the universe is filled with a cosmological constant, we obtain an Einstein static
universe solution for every $\rho_s$, with the corresponding scale factor being
 \begin{eqnarray}
\label{3-7flat}
&&
a_{s}^{2}=
\frac{ \sqrt{\kappa}}{2\alpha\sqrt{2}\sqrt{\rho_s} }.
\end{eqnarray}
Note that the role of a non-zero quantum gravity deformation parameter $\alpha$ is
crucial in making the above solution non-trivial, since in the limit $\alpha\rightarrow0$
it becomes the aforementioned trivial solution.

\subsection{Non-flat universe ($k\neq0$)}

Let us now investigate the non-flat universe. In this case, the system
(\ref{3-3}),(\ref{3-4}) accepts four solutions, i.e four Critical Points (CP), namely
\begin{eqnarray}
\label{3-15}
&&
\!\!\!\!\!\!\!\!\!\!\!\!\!\!
\mbox{CP~1}:~ \left(\frac{1}{a_{s}^2}\right)_1=0,\\
\label{3-15b}
&&
\!\!\!\!\!\!\!\!\!\!\!\!\!\!
\mbox{CP~2}:
~\left(\frac{1}{a_{s}^2}\right)_2=-\frac{3\alpha^{2} c^{6}k(2-w)}
{\kappa^{2}(1+3w)}\nonumber\\
&&
\ \ \ \ \ \ \ \ \ \ \ \ \ \ \ \ \ \,
\times\!\left\{-1+\sqrt{1+\frac{32}{3}\left[\frac{1+3w}{(2-w)^2}
\right]}
\right\}\!,
\end{eqnarray}
\begin{eqnarray}
\label{3-15c}
&&
\!\!\!\!\!\!\!\!\!\!\!\!\!\!
\mbox{CP~3}:
~\left(\frac{1}{a_{s}^2}\right)_3=-\frac{3\alpha^{2}c^{6}k(2-w)}
{\kappa^{2}(1+3w)}\nonumber\\
&&
\ \ \ \ \ \ \ \ \ \ \ \ \ \ \ \ \ \,
\times\!
\left\{-1-\sqrt{1+\frac{32}{3}\left[\frac{1+3w}{(2-w)^2}
\right]}
\right\}\!,\\
&&
\!\!\!\!\!\!\!\!\!\!\!\!\!\!
\mbox{CP~4}:~
\left(\frac{1}{a_{s}^2}\right)_4=-2\sqrt{\frac{q_1}{3}}\sinh\left(\frac{q_2}{
3}\right)~,
\label{3-15end}
\end{eqnarray}
with
\begin{eqnarray}
&&q_1=-\frac{12k^2c^{12}\alpha^4}{\kappa^4}\left[\frac{w^2+20w+12}{(1+3w)^2}
\right],\nonumber\\
&&
q_2=\sinh^{-1} \left(\frac{3q_3}{q_4q_1}\right),\nonumber\\
&&
q_3=\frac{36k^{3}c^6\alpha^6}{\kappa^6}\frac{(2-w)}{(1+3w)^2}\left[1-\frac{
4}{9}\frac{(2-w)^2}{1+3w}
\right],
\nonumber\\
&&
q_4=\frac{4kc^2\alpha^2}{\kappa^2(1+3w)}\sqrt{
-w^2-20w-
12}~.
\end{eqnarray}
The Critical Point $1$ is the trivial one with $a_{s}\rightarrow\infty$. The Critical
Point $2$ is physical for $k=+1$, $w>2$ or  $k=-1$, $-\frac{1}{3}<w<2$, with the second
case being the realistic one. The Critical Point $3$ is physical for $k=+1$,
$-\frac{1}{3}<w<2$ or $k=-1$, $ w>2$, with the first case being the realistic one.
Finally, Critical Point $4$ is physical for $-10-2\sqrt{22}<w<-10+2\sqrt{22}$, which
includes the
cosmological constant value $w=-1$. We stress once again the crucial role of the
deformation parameter $\alpha$ that arises from the generalized uncertainty principle, in
the existence of the three non-trivial critical points, since in the limit
$\alpha\rightarrow0$ all points coincide with the first, trivial one.

For each one of the above solutions for $a_s$, the corresponding $\rho_s$ is given by
\begin{equation}
\label{3-12}
\rho_{s}^{1/2}=-\frac{1}{3A}\left(\kappa+\Delta_2+\frac{\Delta_{0}}{\Delta_2}\right)~,
\end{equation}
where
\begin{eqnarray}
\label{3-12a}
&&
\Delta_2=\sqrt[3]{\frac{\Delta_{1}+\sqrt{\Delta_{1}^{2}-4
\Delta_{0}^{3}}}{2}},\nonumber\\
&&
\Delta_{0}\equiv\kappa^{2}
-3AB,
\nonumber\\
&&
\Delta_{1}\equiv 2\kappa^{3}-9\kappa AB+27A^{2}C~,
\end{eqnarray}
with
\begin{equation}\label{3-12b}
A\equiv-2\sqrt{\frac{8\kappa}{c^4}}\alpha a_{s}^{2},~~~ B\equiv\frac{6\alpha
kc^{4}}{\sqrt{2\kappa}},~~~C\equiv-\frac{kc^{4}}{a_{s}^{2}}.
\end{equation}
 Relation (\ref{3-12}) provides the energy
density $\rho_s$ corresponding to the derived critical points in the static configuration
in terms of $k, \kappa, w, c$ and $\alpha$. In order to examine whether the
resulting values are compatible with the energy of the inflationary epoch, in the
following Table we present the approximate obtained values of $\rho_s$. As we can see, we
acquire a very good compatibility with the energy density at the inflationary phase.
\begin{table}[ht]
	\begin{center}
\begin{tabular}{|c|c|c|}
  \hline
  Number of CP & $(\rho_s)_{k=+1}~kg/m^3$ & $(\rho_s)_{k=-1}~kg/m^3$ \\ \hline
  \mbox{CP 2}& $\sim10^{97}$ & $\sim10^{97}$ \\ \hline
  \mbox{CP 3}& $\sim10^{110}$ & $\sim10^{97}$ \\ \hline
  \mbox{CP 4}& $\sim10^{95}-10^{96}$&$ \sim10^{95}-10^{96}$ \\
  \hline
\end{tabular}
	\end{center}
	\caption{The approximate values of $\rho_s$ corresponding to the
non-trivial obtained
critical points, calculated through (\ref{3-12}). }
	\end{table}

\section{Dynamical stability}
\label{dynamicalanalysis2}

In the previous section we showed that Einstein static universe can be a solution of the
cosmological equations in the framework of the generalized
uncertainty principle. In the present section we desire to analyze the dynamical
stability of these solutions, i.e to see whether the universe
can remain in such a phase for very large time intervals.  

 In order to perform the dynamical analysis one usually
expresses the
cosmological equations as a dynamical system, and performing linear perturbations
around the previously obtained solutions he proceeds to a detailed
phase-space analysis, by examining the eigenvalues of the involved perturbation matrix,
which reveals whether these solutions are stable or unstable \cite{Copeland:1997et}. In
the following we will follow the alternative but equivalent (in cases of 2D
equation systems) approach
of
\cite{Wu:2011xa,Atazadeh:2014xsa,Atazadeh:2015zma,Board:2017ign}.  In particular, we
perturb linearly the
Friedmann equations  (\ref{2-1}) and  (\ref{2-3})  in the regime
$\rho\gg\frac{|k|c^{2}}{\kappa}$, around the obtained Einstein static universe solutions
(\ref{3-7flat}) and  (\ref{3-15})-(\ref{3-15end}).
The perturbations in the scale factor
and matter density read as:
\begin{eqnarray}
\label{4-1}
&&a(t)\rightarrow a_{s}(1+\delta a(t)),\nonumber\\
&&\rho(t)\rightarrow \rho_{s}(1+\delta \rho(t))~.
\end{eqnarray}
Inserting into the first Friedmann equation (\ref{2-1}), using
\begin{eqnarray}
\label{4-2}
&&(1+\delta a(t))^{n}\simeq 1+n\delta a(t),\nonumber\\
&&(1+\delta \rho(t))^{n}\simeq 1+n\delta \rho(t),
\end{eqnarray}
 and neglecting terms with two differentials, we obtain
\begin{eqnarray}
\label{4-3}
&&
\!\!\!\!\!\!\!\!\!\!\!\! \!\!\!\!
\kappa \rho_{s}(1+ \delta
\rho)
-kc^{2}a_{s}^{-2}
-2\sqrt{2\kappa}\alpha a_{s}^{2}\rho_{s}^{3/2}
+2kc^{2}a_{s}^{-2}
\delta a
\nonumber\\
&&
\!\!\!\!\!\!\!\!\!\!\!\! \!\!\!\!
-
\frac{3\alpha kc^{2}}{\sqrt{2\kappa}}\rho_{s}^{1/2}(2+\delta \rho)
-\sqrt{2\kappa}\alpha a_{s}^{2}\rho_{s}^{3/2}\left(3\delta \rho +4\delta a \right)=0.
\end{eqnarray}
Similarly, using (\ref{4-1}) to perturb the second Friedmann equation (\ref{2-3}), and
neglecting terms with two differentials, we obtain
\begin{eqnarray}
\label{4-9}
&&\!\!\!\!\!\!\!\!\!\!\!\!\!\!
\delta \ddot{a}=\left(4\sqrt{2\kappa}\alpha a_{s}^{2}\rho_{s}^{3/2}+\frac{126\alpha
k^{2}c^{4}}
{\sqrt{(2\kappa})^3}a_{s}^{-2}\rho_{s}^{-1/2}\right)\delta a
\nonumber\\
&&
-
\bigg[3\sqrt{2\kappa}\alpha
a_{s}^{2}
\rho_{s}^{3/2}-\frac{3\alpha kc^{2}}{\sqrt{2\kappa}}(2-w)\rho_{s}^{1/2}
 \nonumber\\
&& \ \ \ \,
+\frac{3\alpha
k^{2}c^{4}}{\sqrt{(2\kappa})^3}a_{s}^{-2}\rho_{s}^{-1/2}+\frac{\kappa}{2}(3w+1)\rho_{
s}
\bigg]\delta \rho~.
\end{eqnarray}

In the following two subsections we examine the flat and non-flat cases separately.

\subsection{Flat universe ($k=0$)}

In the case of a flat universe,   (\ref{4-3}) leads to
\begin{eqnarray}
\label{4-4flat}
\left(\frac{\delta \rho}{\delta a}\right) = \left(\frac{4\sqrt{2\kappa}
\alpha a_{s}^{2}\rho_{s}^{3/2}}{\kappa \rho_{s}-3\sqrt{2\kappa}\alpha
a_{s}^{2}\rho_{s}^{3/2}}\right).
\end{eqnarray}
Thus,  inserting  (\ref{4-4flat}) into (\ref{4-9}) and neglecting  terms
higher than ${\cal{O}}(\alpha^{2})$, we acquire
\begin{eqnarray}
\label{4-10flat}
\delta
\ddot{a}+\gamma_{f}
\delta a=0~,
\end{eqnarray}
with
\begin{eqnarray}
\label{4-12}
\gamma_{f}=\frac{6\sqrt{2} \alpha \kappa^{3/2} a_{s}^{2}\rho_{s}^{3/2}
(w+1)}{\kappa  -3\sqrt{2}\alpha a_{s}^2\rho_{s}^{1/2}}.
\end{eqnarray}
However, as we found in subsection \ref{dynanalflat}, the non-trivial Einstein static
solution (\ref{3-7flat}) exists only for $w=-1$, which leads to the limiting value
$\gamma_{f}=0$. Hence, we deduce that the scenario at is stable.

In order to provide an additional verification of the above result, we apply the
procedure of \cite{Huang:2015zma,Wu:2009ah,Bag:2014tta}. We introduce two variables,
namely
$x_{1}=a$ and $x_{2}=\dot{a}$, and hence the linear perturbations of the Friedmann
equation (\ref{2-3}), around the critical point (\ref{3-7flat}) and with $w=-1$, leads to
 \begin{eqnarray}
 \label{3-5}
&&\dot{x}_1=x_2\equiv O_1(x_1,x_2),\\
&&\dot{x}_2=
 \kappa
\rho_s x_1-2\sqrt{2\kappa}\alpha \rho_s^{3/2} x_{1}^{3}
\equiv O_2(x_1,x_2).
\end{eqnarray}
Hence, the eigenvalues square  $\lambda^2$  of the Jacobian matrix
\begin{eqnarray}\label{3-6}
J\bigg(O_1(x_1,x_2),O_2(x_1,x_2)\bigg)=\left(
 \begin{array}{cc}
 \frac{\partial O_1}{\partial x_1} & \frac{\partial O_1}{\partial x_2}  \\
 \frac{\partial O_2}{\partial x_1} & \frac{\partial O_2}{\partial x_2}  \\
 \end{array}
  \right),
  \end{eqnarray}
  calculated at the critical point (\ref{3-7flat}),
 is just
  \begin{equation}\label{3-9}
\lambda^{2}= -2\kappa \rho_s.
\end{equation}
 As we observe, the above eigenvalues square is negative
for all physical cases ($\rho_s>0$), independently of the value of $\alpha$. Thus, the
Einstein static universe in the flat geometry is always stable, as we also found through
(\ref{4-12}). Nevertheless, as we
mentioned above, we note that the presence of a non-zero $\alpha$ still has the crucial
effect to make this critical point non-trivial. 
 \begin{figure}[ht]
\begin{tabular}{c}
\epsfig{figure=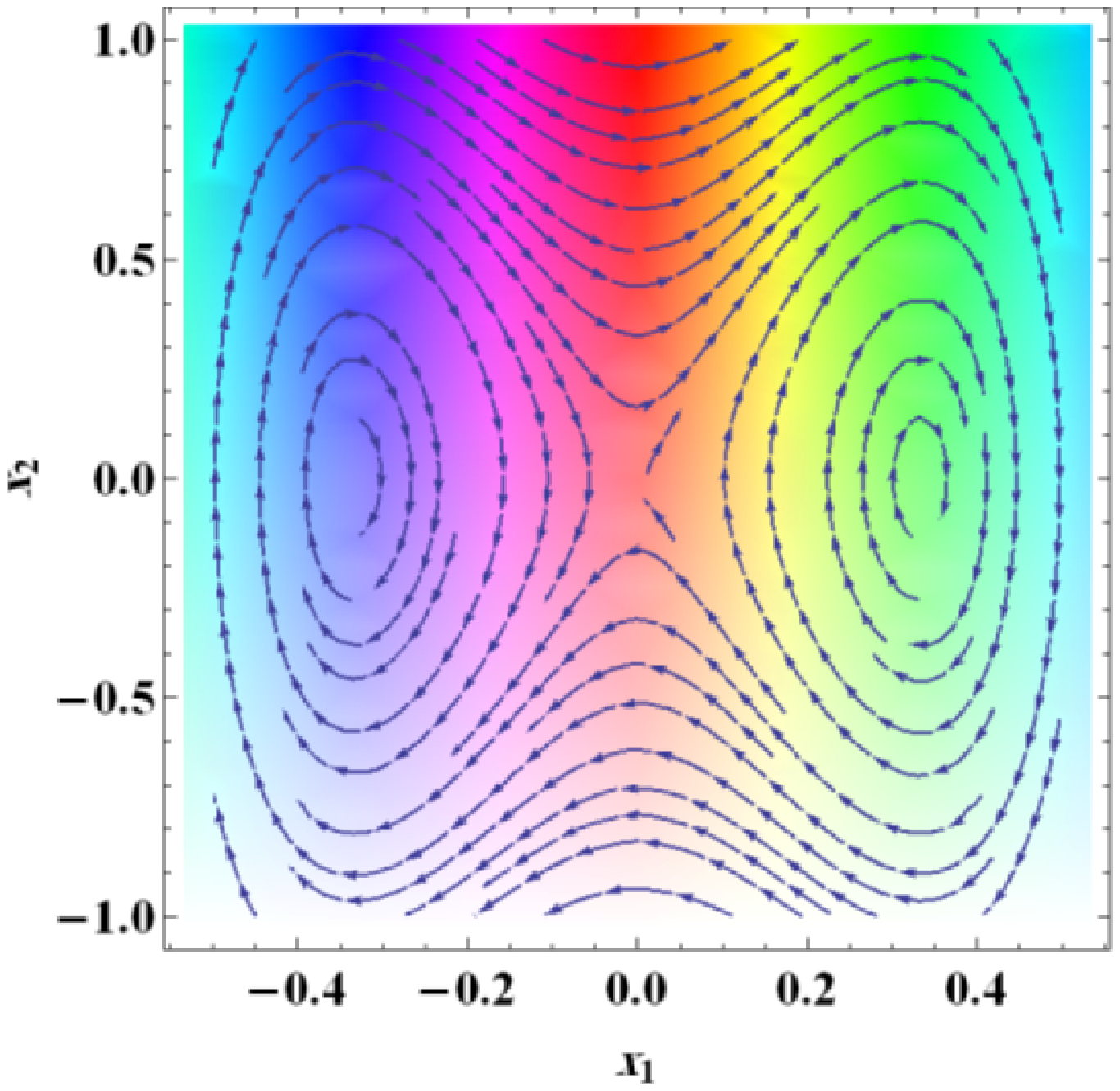,width=6.5cm}\\
\epsfig{figure=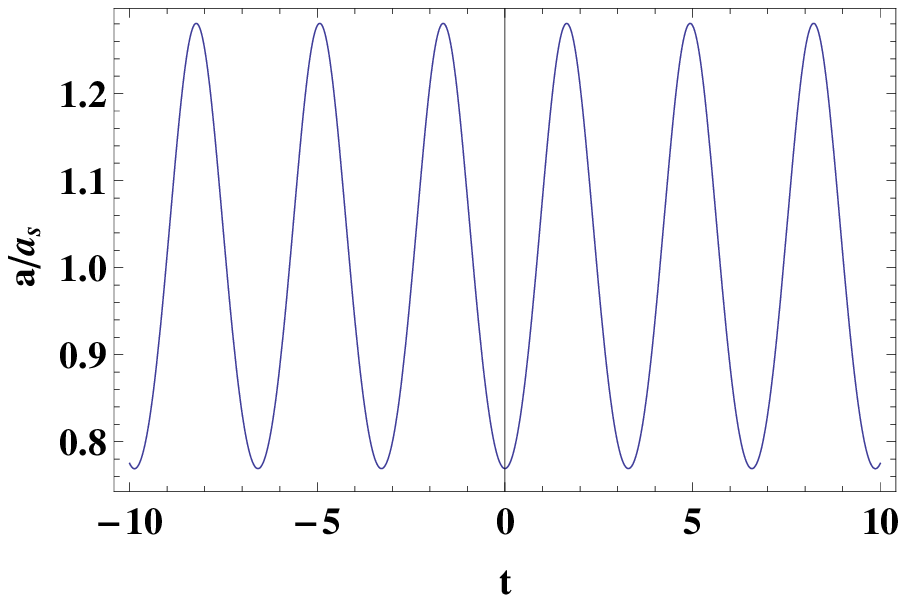,width=6.5cm}
\end{tabular}
\caption{{\it{ The phase diagram in  $(a,\dot{a})$ or $(x_1,x_2)$
space (upper graph) and the evolution of the scale factor (lower graph), for the
spatially flat cosmology, with equation-of-state
parameter $w=-1$, for the choice  ${\alpha}=1$,  in units where  $c=\ell_{pl}=1,
\kappa=1/3$.
The value of $\rho_s$ has been chosen as  $\rho_s=10^{2}$, in order to be consistent with
the condition $\rho\gg\frac{|k|c^{2}}{ \kappa}$. }}}
\label{fig0}
\end{figure}

In order to see the above effect more transparently, in the upper graph of Fig.
\ref{fig0} we present the phase-space behavior  for the spatially flat cosmology with
equation-of-state parameter $w=-1$, where the stable Einstein static universe
critical point is clear (the physical critical point is the one with $x_1>0$).  Moreover,
in order to verify the stability of the Einstein static universe solution in an
alternative way, in the lower graph of Fig. \ref{fig0} we depict the evolution of the
scale factor after a small perturbation around this solution.  As
can be seen, the universe exhibits small oscillations around the Einstein static universe,
without deviating from it, as expected.

In summary, as we can see from the simple case of flat geometry, the effect of the
quantum gravity deformation parameter that arise from the generalized uncertainty
principle is twofold: Firstly, it leads to a non-trivial Einstein static universe
solution that is absent in standard cosmological models, and secondly it leads to its
stabilization. Note that although in some emergent universe scenarios quantum effects are
responsible for destabilization \cite{Mithani:2012ii}, the present incorporation of
quantum effects through natural UV cutoffs is the cause of stabilization. This is one of
the main results of the present work, and will become more
transparent in the more interesting solutions in the case of non-flat geometry.

\subsection{Non-flat universe ($k\neq0$)}

In the case of a non-flat universe,   (\ref{4-3}) leads to
\begin{eqnarray}
\label{4-4}
\left(\frac{\delta \rho}{\delta a}\right) =\left(\frac{4\sqrt{2\kappa}
\alpha a_{s}^{2}\rho_{s}^{3/2}-2 k c^{2}a_{s}^{-2}}{\kappa \rho_{s}-\frac{3\alpha
kc^{2}}{\sqrt{2\kappa}}\rho_{s}^{1/2} -3\sqrt{2\kappa}\alpha
a_{s}^{2}\rho_{s}^{3/2}}\right).
\end{eqnarray}
As expected, in the limit $\alpha\rightarrow0$ we re-obtain the standard result, namely
$\left(\frac{\delta \rho}{\delta a}\right)=-\frac{2 k c^{2}}{a_{s}^{2}\kappa
\rho_{s}}$.
Replacing  (\ref{4-4}) into (\ref{4-9}) and neglecting terms
higher than ${\cal{O}}(\alpha^{2})$, we acquire
\begin{eqnarray}
\label{4-10}
\delta
\ddot{a}+\gamma_{nf}
\delta a=0~,
\end{eqnarray}
with
\begin{eqnarray}
\label{4-12b}
\gamma_{nf}=\left[\frac{\kappa\rho_{s}(f_{1}+f_{2})+f_{3}+f_{4}+f_{5}+f_{6}+f_{7}}
{f_{8}}\right]~,
\end{eqnarray}
and
\begin{eqnarray}\label{11}
&&f_{1}= 4\sqrt{2\kappa}\alpha a_{s}^{2}\rho_{s}^{3/2},\nonumber\\
&&
f_{2}=\frac{12\alpha k^{2}c^{4}}{\sqrt{8\kappa^3}}a_{s}^{-2}\rho_{s}^{-1/2},\nonumber\\
&&
f_{3}=2\sqrt{2\kappa^{3}}\alpha (3w+1) a_{s}^{2}\rho_{s}^{5/2},\nonumber\\
&&
f_{4}=-6\sqrt{2\kappa}\alpha k c^{2} \rho_{s}^{3/2},\nonumber\\
&&
f_{5}=-\frac{6\alpha k^{3} c^{6}}{\sqrt{8\kappa^{3}\rho_{s}}a_{s}^{4}},\nonumber\\
&&
f_{6}=\frac{6\alpha k^{2} c^{4}\rho_{s}^{1/2}(2-w)}{\sqrt{2\kappa}a_{s}^2},\nonumber\\
&&
f_{7}=-kc^{2}\kappa(3w+1)\rho_{s}a_{s}^{-2},\nonumber\\
&&
f_{8}=-\frac{3}{4}f_{1}+\frac{3\alpha
kc^{2}}{\sqrt{2\kappa}}\rho_{s}^{1/2}+\kappa\rho_{s},
 \end{eqnarray}
 and where $a_s$ and $\rho_s$ have to be replaced from  (\ref{3-15})-(\ref{3-15end}) for
the four Einstein static universes respectively. Equation (\ref{4-10}) is the perturbation
equation of FRW cosmology in the case of generalized uncertainty principle. As expected,
in the limit $\alpha\rightarrow0$ it reduces to the standard result, namely
\begin{eqnarray}
\label{4-11}
\delta \ddot{a}- \frac{kc^{2}}{a_{s}^{2}}(3w+1)\delta a_{s}=0~.
\end{eqnarray}

Let us now examine whether   we can obtain $\gamma_{nf}>0$. The form of
$\gamma_{nf}$ given in (\ref{4-12b}), calculated at the non-trivial  critical points
(\ref{3-15b})-(\ref{3-15end}), is too complicated to accept any analytical
treatment, and thus we will examine the value of  $\gamma_{nf}$ numerically.
 \begin{figure}[ht]
 \epsfig{figure=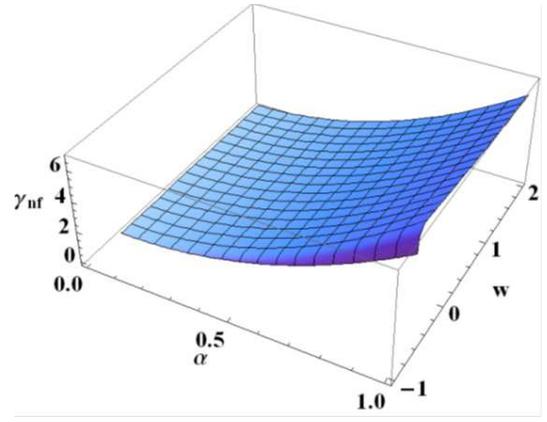,width=7cm}
\caption{{\it{  The coefficient of the perturbation equation
$\gamma_{nf}$ versus the quantum gravity deformation parameter $\alpha$
and the equation-of-state parameter $w$, for the case of
Critical Point $2$ of (\ref{3-15b}),  in the case of open geometry,  in units where
$c=\ell_{pl}=1, \kappa=1/3$. }}}
\label{gammasnonflatpoint2}
\end{figure}
  \begin{figure}[ht]
   \epsfig{figure=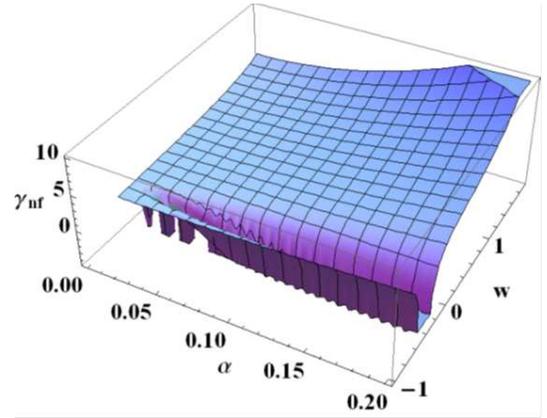,width=7cm}
\caption{{\it{  The coefficient of the perturbation equation
$\gamma_{nf}$ versus the quantum gravity deformation parameter $\alpha$
and the equation-of-state parameter $w$, for the case of
Critical Point $3$ of (\ref{3-15c}),  in the case of closed geometry,  in units where
$c=\ell_{pl}=1, \kappa=1/3$. }}}
\label{gammasnonflatpoint3}
\end{figure}
 In
Fig. \ref{gammasnonflatpoint2} we present $\gamma_{nf}$ versus $\alpha$ and $w$, for the
case of Critical Point $2$ of (\ref{3-15b}), in the case of open geometry (since for the
open geometry this point exists for the more physically interesting $w$-interval, namely
$-\frac{1}{3}<w<2$).
As we can see, for the regions of its existence,  we obtain $\gamma_{nf}>0$ if $\alpha$
acquires positive values. Hence, the scenario at hand can be stable.
Similarly, in Fig. \ref{gammasnonflatpoint3} we present $\gamma_{nf}$ versus $\alpha$ and
$w$, for the Critical Point $3$ of (\ref{3-15c}), in the case of closed geometry
(where $-\frac{1}{3}<w<2$). As we observe, this point can be stable for suitable choices
of $\alpha$ and $w$.
Finally, in Fig. \ref{gammasnonflatpoint4} we present $\gamma_{nf}$ versus $\alpha$
and $w$, for the  Critical Point $4$ of (\ref{3-15end}), for a part of the range of
its existence, namely for $-10-2\sqrt{22}<w<-10+2\sqrt{22}$,  in the case
of closed and open geometry. Similarly to the previous critical points, we can see that
for suitable values of $\alpha$ and $w$ this point, which is the most interesting one
concerning the successful realization of the emergent universe scenario, is stable.
\begin{figure}[ht]
  \epsfig{figure=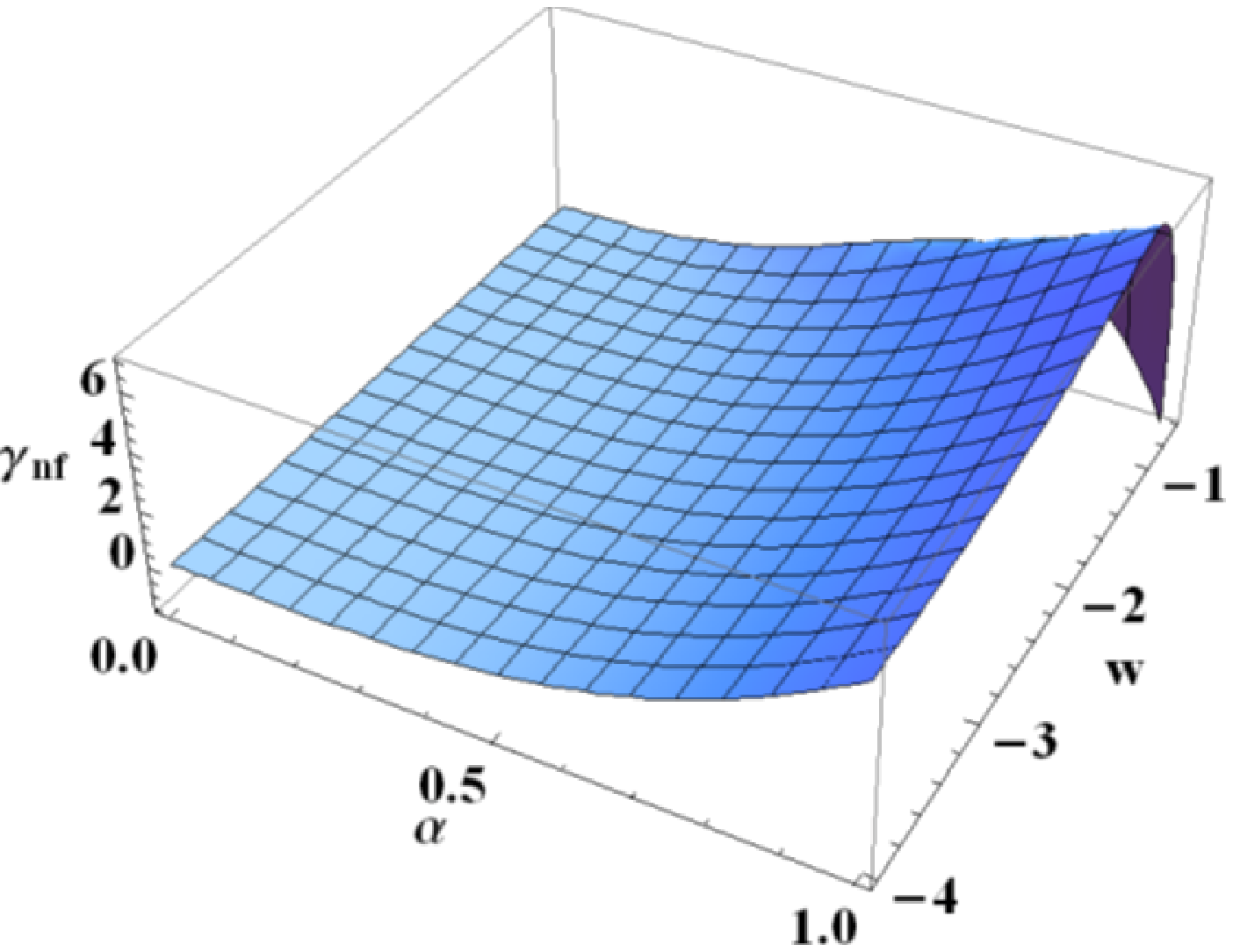,width=7cm}\\
     \epsfig{figure=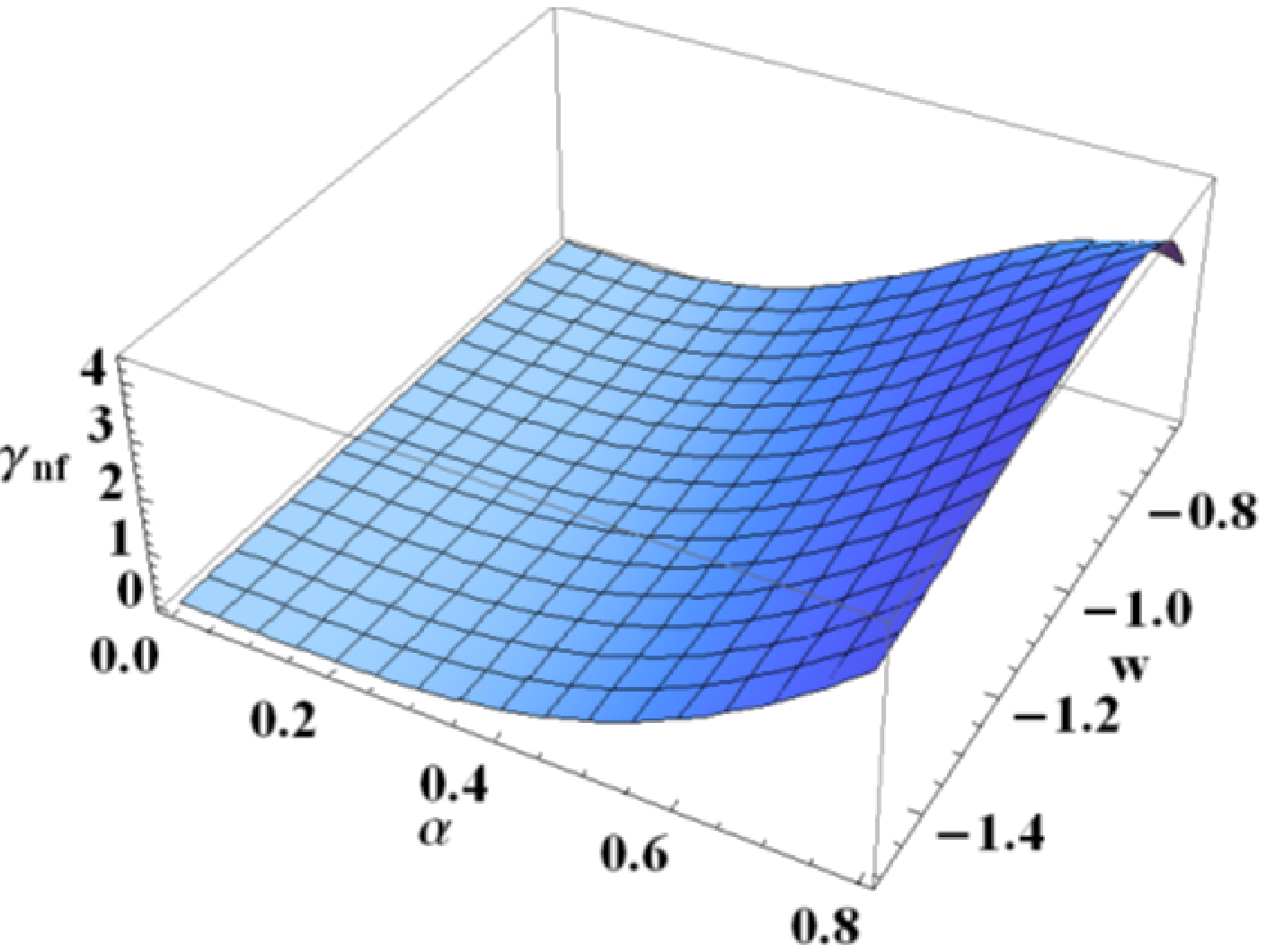,width=7cm}
\caption{{\it{  The coefficient of the perturbation equation
$\gamma_{nf}$ versus the quantum gravity deformation parameter $\alpha$
and the equation-of-state parameter $w$, for the case of
Critical Point $4$ of (\ref{3-15end}),  in the case of closed  (upper graph) and open
(lower graph) geometry,  in units where
$c=\ell_{pl}=1, \kappa=1/3$. }}}
\label{gammasnonflatpoint4}
\end{figure}


Let us verify the above results using the approach of
\cite{Huang:2015zma,Wu:2009ah,Bag:2014tta}, and express them in a more transparent way. We
introduce the two variables
$x_{1}=a$ and $x_{2}=\dot{a}$, and therefore the linear perturbations of the Friedmann
equation (\ref{2-3}) around the Critical Points (\ref{3-15})-(\ref{3-15end})
leads to
\begin{eqnarray}
 \label{3-5}
&&\!\!\!\!\!\!\!\!\!\!\!\!\!
\dot{x}_1=x_2\equiv O_1(x_1,x_2),\\
&&\!\!\!\!\!\!\!\!\!\!\!\!\!
\dot{x}_2=\left[
\frac{3kc^2}{2\sqrt{2\kappa}}(2-w) \alpha
-\frac{\kappa}{2}(1+3w)
\rho_s\right]x_1\nonumber\\
&&
-2\sqrt{2\kappa}\alpha \rho_s^{3/2} x_{1}^{3}+
\frac{3k^{2}c^{4}\alpha}
{\sqrt{2\kappa^3\rho_s}}
\frac{1}{x_{1}}
\equiv O_2(x_1,x_2).
\end{eqnarray}
The eigenvalues square of the Jacobian matrix (\ref{3-6}), for the above
$O_1(x_1,x_2)$,$O_2(x_1,x_2)$, calculated at the non-trivial Critical Points
$2$,~$3$ and $4$, read as follows.

For the Critical Points  $2$ and $3$ we have
\begin{eqnarray}
\label{3-18}
&&\!\!\!\!\!\!
\lambda^{2}=-\frac{9\alpha k^{2}c^{5}}{16\kappa}\left[\frac{(2-w)^2}{1+3w
}\right]\left[
2f(w)_{\pm}^2+f(w)_{\pm}\right]
\nonumber\\
&&\ \ \ \
+\frac{2\kappa^4}{9\alpha^{4}kc^2}\left(\frac{
1+3w}{2-
w}\right)^3
f(w)_{\pm}^{-3}~,
\end{eqnarray}
with
$ f(w)_{\pm}\equiv
-1\pm\sqrt{1+\frac{32}{3}\left[\frac{1+3w}{(2-w)^2}\right]
}$, where the plus sign corresponds to Critical Point $2$
and the minus sign to Critical Point $3$. As we can see, for the range where they are
physical, namely for $k=+1$, $w>2$ or  $k=-1$, $-\frac{1}{3}<w<2$ for Critical Point $2$,
and
for  $k=+1$, $-\frac{1}{3}<w<2$ or $k=-1$, $ w>2$  for Critical Point $3$, we always get
$\lambda^2<0$, and thus both points are always
stable.

\begin{figure}[ht]
\epsfig{figure=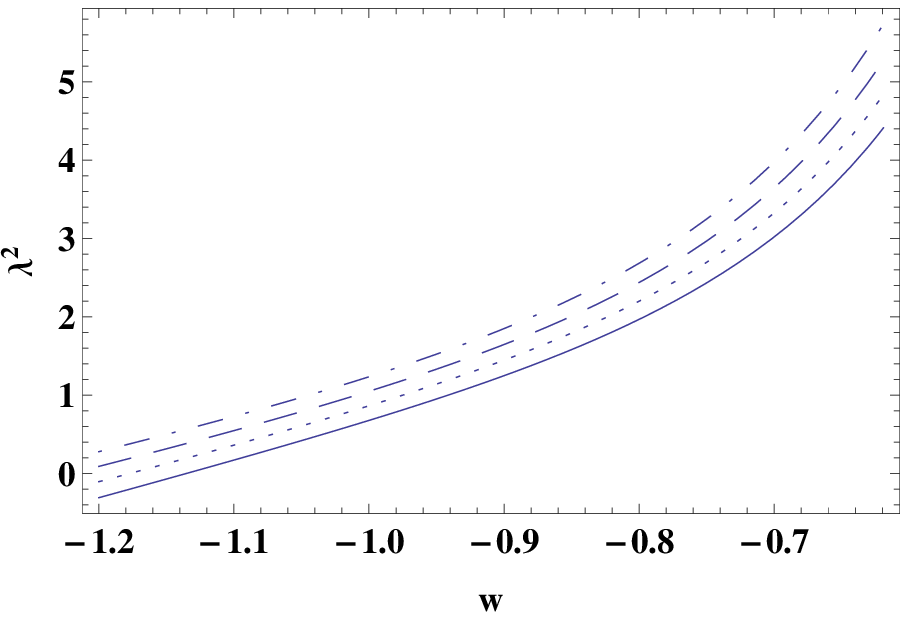,width=7.7cm}\\
\hspace{-0.55cm}
\epsfig{figure=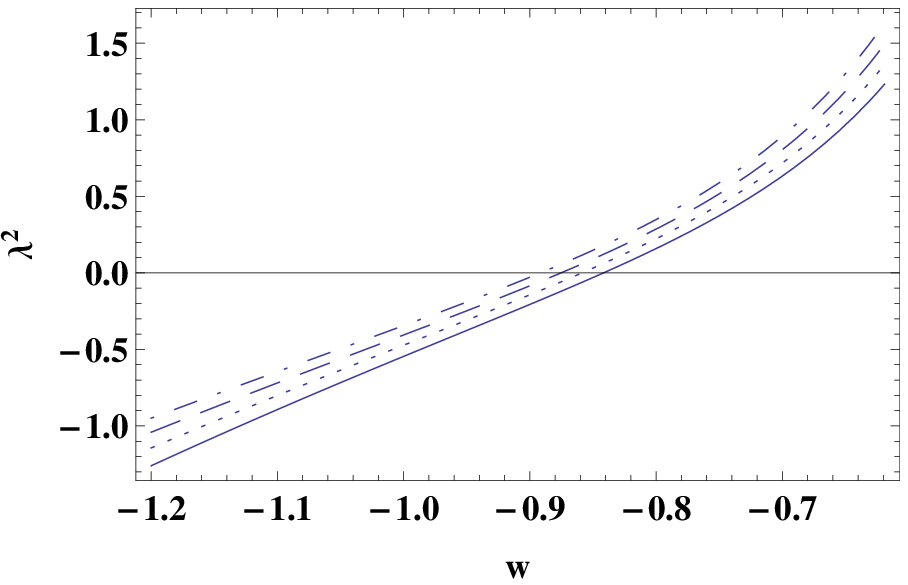,width=8.25cm}
\caption{\it{The eigenvalues square  $\lambda^2$ versus the equation-of-state
parameter $w$ for the critical point $4$ given in (\ref{3-15end}), for closed (upper
graph) and open (lower graph) geometry, for various values of $\alpha$, namely
$\alpha=0.88$ (solid curve) $\alpha=0.92$ (dotted curve)  $\alpha= 0.96$ (dashed curve)
$\alpha=1$ (dashed-dot curve), in units where  $c=\ell_{pl}=1, \kappa=1/3$.} }
\label{fig2}
\end{figure}

Let us now examine the stability of the Critical Point $4$ given in (\ref{3-15end}). The
corresponding eigenvalues square of the Jacobian matrix is found to be
\begin{eqnarray}
\label{3-20}
&&\!\!\!\!\!\!\!\!
\lambda^{2}=-\frac{9\alpha^{2}}{2\sqrt{2}\kappa^2q_1\sinh^{2}(\frac{q_2}{3})}
+\frac{3kc^2(w-2)}{16}\sqrt{\frac{q_1}{3}}
\sinh\left(\frac{q_2}{3}\right)\nonumber\\
&&\ \ \
-\frac{\kappa^2q_1}{12\alpha^2}(1+3w)\sinh^{2}
\left(\frac
{q_2}{3}\right)~.
\end{eqnarray}
Hence, we deduce that the sign of $\lambda^2$ depends on both $\alpha$ and $w$ in a
complicated way that does not allow for an analytical treatment, and thus in order to
examine its behavior we will resort to numerical elaboration. 
In   Fig. \ref{fig2} we depict
$\lambda^2$ versus $w$ for various values of $\alpha$, for the case of closed and open
geometry  (as we mentioned earlier we consider the
realistic values of $\tilde{\alpha}$, and thus of $\alpha$ in the units we use, to be
around 1  in
order for minimal length effects to be important only around the Planck length and not
introduce a new physical scale between the Planck and the electroweak scale
\cite{Ali:2011fa,Marin:2013pga,Das:2008kaa}).

The Critical Point 4 exhibits a very interesting behavior concerning the realization of 
the exit from the static universe and of the emergent universe
scenario. In particular, for both spatial geometries we can have a
stable Einstein static universe (Critical Point 2 for $k=-1$ or Critical Point 3 for 
$k=+1$) for very long time intervals (infinite in the past if $w$ approaches $0$ in the 
past).
As time passes and the universe equation-of-state parameter decreases these critical 
points become unstable and are exchanged with their unstable counterpart Critical Point 
4, offering a natural graceful exit from the Einstein static universe and an entering 
into the usual expanding thermal history (this procedure is more efficient for the 
closed geometry, since Critical Point 4 is always unstable for suitable values  of 
$\alpha$).    
This behavior is also achieved in complicated models
of the emergent universe in some various modified gravities \cite{Huang:2015zma}, however
in
the present scenario it is obtained solely from the quantum gravity deformation parameter
$\alpha$.

Hence, from the above we can deduce the central role of the
quantum gravity deformation parameter that arises from the generalized uncertainty
principle: It leads to non-trivial Einstein static universe
solutions that are absent in standard cosmological models, and it provides a
mechanism for a successful exit from a stable Einstein static universe into the expanding
thermal history, i.e. for a complete realization of the emergent universe scenario. This
is one of the main results of the present work.

\section{Discussions and conclusions}
\label{Conclusions}

In the present work we investigated the realization of the emergent universe scenario in
the framework of theories with natural UV cutoffs. In particular, we considered the
generalized uncertainty principle, which includes a deformation parameter $\alpha$ that
arises from quantum gravity modifications corresponding to two natural cutoffs, namely a
minimum length and a maximum momentum. Applying it in a cosmological framework we
obtained the modified Friedmann equations and we studied them in detail to see whether we
can acquire Einstein static universe solutions, which are the basic concept in the
realization of the emergent universe scenario.

As a first step we extracted the Einstein static universe solutions, analyzing for
convenience the flat and non-flat cases separately.
Then, performing a dynamical analysis in the phase space we examined the
dynamical stability of these solutions. As we showed, the role of the new deformation
parameter $\alpha$ is crucial in a twofold way. Firstly, it leads to the appearance of
new Einstein static universe critical points, that are absent in standard cosmology, and
this is true also in the flat case where standard cosmology does not accept an Einstein
static universe. Secondly, the deformation parameter $\alpha$ plays a central role in
providing a mechanism for a graceful exit from a stable Einstein static universe into
the expanding thermal history, i.e. for a complete and successful realization of the
emergent universe scenario. This double role of $\alpha$, i.e. of the quantum gravity
modifications arising from the natural UV cutoffs, is one of the main results of the
present work.

In summary, we conclude that the emergent universe scenario can be successfully realized
in the framework of cosmology with generalized uncertainty principle arising from the
presence of natural UV cutoffs.

\begin{acknowledgements}
The work of KN has been financially supported  by the Center for Excellence in
Astronomy and Astrophysics of IRAN (CEAAI-RIAAM) under research project No. 1/4717-70.
This article is based upon work from COST Action ``Cosmology and Astrophysics Network
for Theoretical Advances and Training Actions'', supported by COST (European Cooperation
in Science and Technology).
\end{acknowledgements}


\begin{thebibliography}{99}


\bibitem{Guth:1980zm}
  A.~H.~Guth,
  Phys.\ Rev.\ D {\bf 23}, 347 (1981);
  K.~Sato,
  Mon.\ Not.\ Roy.\ Astron.\ Soc.\  {\bf 195}, 467 (1981).
  A.~D.~Linde,
  Phys.\ Lett.\  {\bf 108B}, 389 (1982).

\bibitem{Borde:2001nh}
  A.~Borde, A.~H.~Guth and A.~Vilenkin,
  Phys.\ Rev.\ Lett.\  {\bf 90}, 151301 (2003).

\bibitem{Mukhanov:1991zn}
  V.~F.~Mukhanov and R.~H.~Brandenberger,
  Phys.\ Rev.\ Lett.\  {\bf 68}, 1969 (1992).



\bibitem{Novello:2008ra}
  M.~Novello and S.~E.~P.~Bergliaffa,
  Phys.\ Rept.\  {\bf 463}, 127 (2008).



\bibitem{Ellis:2002we}
  G.~F.~R.~Ellis and R.~Maartens,
  Class.\ Quant.\ Grav.\  {\bf 21}, 223 (2004);
  G.~F.~R.~Ellis, J.~Murugan and C.~G.~Tsagas,
  Class.\ Quant.\ Grav.\  {\bf 21}, no. 1, 233 (2004).





\bibitem{Eddington}
A. S. Eddington,  Mon. Not. Roy. Astron. Soc.  \textbf{90}, 668 (1930);
G. W. Gibbons,   Nucl. Phys. B \textbf{310}, 636 (1988).



\bibitem{Copeland:2006wr}
  E.~J.~Copeland, M.~Sami and S.~Tsujikawa,
  Int.\ J.\ Mod.\ Phys.\  D {\bf 15}, 1753 (2006).




\bibitem{Cai:2009zp}
  Y.~F.~Cai, E.~N.~Saridakis, M.~R.~Setare and J.~Q.~Xia,
  Phys.\ Rept.\  {\bf 493}, 1 (2010).



\bibitem{Nojiri:2006ri}
  S.~Nojiri and S.~D.~Odintsov,
  eConf {\bf C0602061}, 06 (2006), Int.\ J.\ Geom.\ Meth.\ Mod.\ Phys.\
{\bf 4}, 115 (2007).


\bibitem{Capozziello:2011et}
  S.~Capozziello and M.~De Laurentis,
  Phys.\ Rept.\  {\bf 509}, 167 (2011).







 \bibitem{Brandenberger:1993ef}
  R.~H.~Brandenberger, V.~F.~Mukhanov and A.~Sornborger,
  Phys.\ Rev.\  D {\bf 48}, 1629 (1993);
  J.~Khoury, B.~A.~Ovrut, N.~Seiberg, P.~J.~Steinhardt and N.~Turok,
  Phys.\ Rev.\ D {\bf 65}, 086007 (2002);
  M.~Bojowald,
  Phys.\ Rev.\ Lett.\  {\bf 86}, 5227 (2001);
  J.~Khoury, B.~A.~Ovrut, P.~J.~Steinhardt and N.~Turok,
  Phys.\ Rev.\  D {\bf 64}, 123522
(2001);
  Y.~Shtanov and V.~Sahni,
  Phys.\ Lett.\  B {\bf 557}, 1 (2003);
  J.~Martin and P.~Peter,
  Phys.\ Rev.\  D {\bf 68}, 103517 (2003);
  P.~Creminelli and L.~Senatore,
  JCAP {\bf 0711}, 010 (2007);
  E.~N.~Saridakis,
  Nucl.\ Phys.\ B {\bf 808}, 224 (2009);
  J.~L.~Lehners,
  Phys.\ Rept.\  {\bf 465}, 223 (2008);
R.~Brandenberger,
  Phys.\ Rev.\  D {\bf 80}, 043516 (2009);
  Y.~-F.~Cai and E.~N.~Saridakis,
  Class.\ Quant.\ Grav.\  {\bf 28}, 035010 (2011);
  Y.~F.~Cai, D.~A.~Easson and R.~Brandenberger,
  JCAP {\bf 1208}, 020 (2012);
  T.~Biswas, A.~S.~Koshelev, A.~Mazumdar and S.~Y.~Vernov,
  JCAP {\bf 1208}, 024 (2012);
  Y.~F.~Cai, C.~Gao and E.~N.~Saridakis,
  JCAP {\bf 1210} (2012) 048;
  T.~Qiu, X.~Gao and E.~N.~Saridakis,
  Phys.\ Rev.\ D {\bf 88}, no. 4, 043525 (2013);
  Y.~F.~Cai, J.~Quintin, E.~N.~Saridakis and E.~Wilson-Ewing,
  JCAP {\bf 1407}, 033 (2014).


\bibitem{Boehmer:2003iv}
  C.~G.~Boehmer,
  Class.\ Quant.\ Grav.\  {\bf 21}, 1119 (2004);
  K.~Atazadeh,
  JCAP {\bf 1406}, 020 (2014).




\bibitem{Barrow:1983rx}
  J.~D.~Barrow and A.~C.~Ottewill,
  J.\ Phys.\ A {\bf 16}, 2757 (1983);
  C.~G.~Boehmer, L.~Hollenstein and F.~S.~N.~Lobo,
  Phys.\ Rev.\ D {\bf 76}, 084005 (2007);
  R.~Goswami, N.~Goheer and P.~K.~S.~Dunsby,
  Phys.\ Rev.\ D {\bf 78}, 044011 (2008).

\bibitem{Wu:2011xa}
  P.~Wu and H.~Yu,
  Phys.\ Lett.\ B {\bf 703}, 223 (2011);
  J.~T.~Li, C.~C.~Lee and C.~Q.~Geng,
  Eur.\ Phys.\ J.\ C {\bf 73}, no. 2, 2315 (2013).


\bibitem{Mulryne:2005ef}
  D.~J.~Mulryne, R.~Tavakol, J.~E.~Lidsey and G.~F.~R.~Ellis,
  Phys.\ Rev.\ D {\bf 71}, 123512 (2005);
  L.~Parisi, M.~Bruni, R.~Maartens and K.~Vandersloot,
  Class.\ Quant.\ Grav.\  {\bf 24}, 6243 (2007).


\bibitem{Parisi:2012cg}
  L.~Parisi, N.~Radicella and G.~Vilasi,
  Phys.\ Rev.\ D {\bf 86}, 024035 (2012);
  K.~Zhang, P.~Wu and H.~Yu,
  Phys.\ Rev.\ D {\bf 87}, no. 6, 063513 (2013).


\bibitem{Khodadi:2015fav}
  M.~Khodadi, Y.~Heydarzade, F.~Darabi and E.~N.~Saridakis,
  Phys.\ Rev.\ D {\bf 93}, no. 12, 124019 (2016);
  Y.~Heydarzade, M.~Khodadi and F.~Darabi,
  arXiv:1502.04445 [gr-qc].



\bibitem{Gergely:2001tn}
  L.~A.~Gergely and R.~Maartens,
  Class.\ Quant.\ Grav.\  {\bf 19}, 213 (2002);
  K.~Zhang, P.~Wu and H.~W.~Yu,
  Phys.\ Lett.\ B {\bf 690}, 229 (2010);

\bibitem{Atazadeh:2014xsa}
  K.~Atazadeh, Y.~Heydarzade and F.~Darabi,
  Phys.\ Lett.\ B {\bf 732}, 223 (2014);
  Y.~Heydarzade and F.~Darabi,
  JCAP {\bf 1504}, no. 04, 028 (2015).



\bibitem{Cai:2012yf}
  Y.~F.~Cai, M.~Li and X.~Zhang,
  Phys.\ Lett.\ B {\bf 718}, 248 (2012);
  Y.~F.~Cai, Y.~Wan and X.~Zhang,
  Phys.\ Lett.\ B {\bf 731}, 217 (2014).








\bibitem{Hossenfelder:2012jw}
  S.~Hossenfelder,
  Living Rev.\ Rel.\  {\bf 16}, 2 (2013).


\bibitem{AmelinoCamelia:2000mn}
  G.~Amelino-Camelia,
  Int.\ J.\ Mod.\ Phys.\ D {\bf 11}, 35 (2002);
  G.~Amelino-Camelia,
  Phys.\ Lett.\ B {\bf 510}, 255 (2001);
  J.~Magueijo and L.~Smolin,
  Class.\ Quant.\ Grav.\  {\bf 21}, 1725 (2004);
  J.~Magueijo and L.~Smolin,
  Phys.\ Rev.\ D {\bf 67}, 044017 (2003).



\bibitem{Amati:1988tn}
  D.~Amati, M.~Ciafaloni and G.~Veneziano,
  Phys.\ Lett.\ B {\bf 216}, 41 (1989).



\bibitem{Ali:2009zq}
  A.~F.~Ali, S.~Das and E.~C.~Vagenas,
  Phys.\ Lett.\ B {\bf 678}, 497 (2009);
  S.~Das, E.~C.~Vagenas and A.~F.~Ali,
  Phys.\ Lett.\ B {\bf 690}, 407 (2010)
  Erratum: [Phys.\ Lett.\ B {\bf 692}, 342 (2010)];
  K.~Nozari and A.~Etemadi,
mechanics,''
  Phys.\ Rev.\ D {\bf 85} 104029 (2012).

\bibitem{Ali:2011fa}
  A.~F.~Ali, S.~Das and E.~C.~Vagenas,
  Phys.\ Rev.\ D {\bf 84}, 044013 (2011).



\bibitem{Marin:2013pga}
  F.~Marin {\it et al.},
  Nature Phys.\  {\bf 9}, 71 (2013).



\bibitem{Das:2008kaa}
  S.~Das and E.~C.~Vagenas,
  Phys.\ Rev.\ Lett.\  {\bf 101}, 221301 (2008);
  P.~Jizba, H.~Kleinert and F.~Scardigli,
  Phys.\ Rev.\ D {\bf 81}, 084030 (2010);
  K.~Nozari and P.~Pedram,
  Europhys.\ Lett.\  {\bf 92}, 50013 (2010);
  P.~Pedram, K.~Nozari and S.~H.~Taheri,
  JHEP {\bf 1103}, 093 (2011);
  P.~Wang, H.~Yang and X.~Zhang,
  Phys.\ Lett.\ B {\bf 718}, 265 (2012);
  I.~Pikovski, M.~R.~Vanner, M.~Aspelmeyer, M.~S.~Kim and C.~Brukner,
  Nature Phys.\  {\bf 8}, 393 (2012);
  F.~Marin {\it et al.},
  Nature Phys.\  {\bf 9}, 71 (2013);
  S.~Jalalzadeh, M.~A.~Gorji and K.~Nozari,
  Gen.\ Rel.\ Grav.\  {\bf 46}, 1632 (2014);
  S.~Ghosh,
  Class.\ Quant.\ Grav.\  {\bf 31}, 025025 (2014);
  K.~Nozari, M.~Khodadi and M.~A.~Gorji,
  Europhys.\ Lett.\  {\bf 112}, no. 6, 60003 (2015);
      F.~Scardigli and R.~Casadio,
  Eur.\ Phys.\ J.\ C {\bf 75}, no. 9, 425 (2015);
  M.~Khodadi,
  Astrophys.\ Space Sci.\  {\bf 358}, no. 2, 45 (2015).



\bibitem{Ali:2014hma}
  A.~F.~Ali and B.~Majumder,
  Class.\ Quant.\ Grav.\  {\bf 31}, no. 21, 215007 (2014).

 \bibitem{Singh:2015jus}
  P.~Singh and S.~K.~Soni,
  Class.\ Quant.\ Grav.\  {\bf 33}, no. 12, 125001 (2016).

 \bibitem{Ashtekar:2003}
   A.~Ashtekar, S.~Fairhurst and J.~L.~Willis,
  Class.\ Quant.\ Grav.\  {\bf 20}, 1031 (2003).


 \bibitem{Hossain:2010}
  G.~M.~Hossain, V.~Husain and S.~S.~Seahra,
  Class.\ Quant.\ Grav.\  {\bf 27}, 165013 (2010).


\bibitem{Ashtekar:2006es}
  A.~Ashtekar, T.~Pawlowski, P.~Singh and K.~Vandersloot,
  Phys.\ Rev.\ D {\bf 75}, 024035 (2007).

 \bibitem{Singh:2010qa}
  P.~Singh and F.~Vidotto,
  Phys.\ Rev.\ D {\bf 83}, 064027 (2011).


 \bibitem{Battisti:2009zzb}
  M.~V.~Battisti,
  Phys.\ Rev.\ D {\bf 79}, 083506 (2009).





\bibitem{Copeland:1997et}
  E.~J.~Copeland, A.~R.~Liddle and D.~Wands,
  Phys.\ Rev.\  D {\bf 57}, 4686 (1998);
  P.~G.~Ferreira and M.~Joyce,
  Phys.\ Rev.\ Lett.\  {\bf 79}, 4740 (1997);
  X.~m.~Chen, Y.~g.~Gong and E.~N.~Saridakis,
  JCAP {\bf 0904}, 001 (2009);
  G.~Leon and E.~N.~Saridakis,
  JCAP {\bf 1303}, 025 (2013).





\bibitem{Atazadeh:2015zma}
  K.~Atazadeh and F.~Darabi,
  Phys.\ Lett.\ B {\bf 744}, 363 (2015).


\bibitem{Board:2017ign}
  C.~V.~R.~Board and J.~D.~Barrow,
  arXiv:1709.09501 [gr-qc].





\bibitem{Huang:2015zma}
  Q.~Huang, P.~Wu and H.~Yu,
  Phys.\ Rev.\ D {\bf 91}, no. 10, 103502 (2015).






\bibitem{Wu:2009ah}
  S.~Carneiro and R.~Tavakol,
  Phys.\ Rev.\ D {\bf 80}, 043528 (2009);
  P.~Wu and H.~W.~Yu,
  Phys.\ Rev.\ D {\bf 81}, 103522 (2010);
  K.~Zhang, P.~Wu and H.~Yu,
  JCAP {\bf 1401}, 048 (2014).


\bibitem{Bag:2014tta}
  S.~Bag, V.~Sahni, Y.~Shtanov and S.~Unnikrishnan,
  JCAP {\bf 1407}, 034 (2014);
  H.~Shabani and A.~H.~Ziaie,
  Eur.\ Phys.\ J.\ C {\bf 77}, no. 1, 31 (2017).








\bibitem{Mithani:2012ii}
  A.~Mithani and A.~Vilenkin,
  arXiv:1204.4658 [hep-th];
  A.~T.~Mithani and A.~Vilenkin,
  JCAP {\bf 1405}, 006 (2014).














\end{thebibliography}
\end{document}